\documentclass[aps,superscriptaddress,nofootinbib,
twocolumn,showpacs,secnumarabic,a4paper,floatfix]{revtex4}

\usepackage{graphicx,setspace}
\usepackage{amsfonts,amsmath,amssymb,bm}
\usepackage[dvips,usenames]{color}

\begin{document}

\title{Pulling and Pushing a Cargo With a Catalytically Active Carrier}


\author{M. N. Popescu}
\email{Mihail.Popescu@unisa.edu.au}
\affiliation{
Ian Wark Research Institute, University of South Australia,
5095 Adelaide, South Australia, Australia}
\affiliation{
Max-Planck-Institut f\"ur Metallforschung, Heisenbergstr. 3,
70569 Stuttgart, Germany}
\author{M. Tasinkevych}
\email{miko@mf.mpg.de}
\affiliation{
Max-Planck-Institut f\"ur Metallforschung, Heisenbergstr. 3,
70569 Stuttgart, Germany}
\affiliation{Institut f\"ur Theoretische und Angewandte Physik,
Universit\"at Stuttgart, Pfaffenwaldring 57, 70569 Stuttgart,
Germany}
\author{S. Dietrich}
\email{dietrich@mf.mpg.de}
\affiliation{
Max-Planck-Institut f\"ur Metallforschung, Heisenbergstr. 3,
70569 Stuttgart, Germany}
\affiliation{Institut f\"ur Theoretische und Angewandte Physik,
Universit\"at Stuttgart, Pfaffenwaldring 57, 70569 Stuttgart,
Germany}

\begin{abstract}
Catalytically active particles suspended in a liquid can move due to 
self-phoresis by generating solute gradients via chemical reactions 
of the solvent occurring at parts of their surface.
Such particles 
can be used as carriers at the micro-scale. As a simple model for a 
carrier-cargo system we consider a catalytically active particle 
connected by a thin rigid rod to a catalytically inert cargo 
particle. 
We show that the velocity of the composite strongly depends 
on the relative orientation of the carrier-cargo link.
Accordingly, there 
is an optimal configuration for the linkage. 
The subtlety of 
such carriers is underscored by the 
observation that a spherical particle completely covered by catalyst, 
which is motionless when isolated, acts as a carrier 
once attached to a cargo.
\end{abstract}


\maketitle

\textbf{\textit{Introduction.}}
\label{sec_intro}
The increasing interest in the development of ``lab on a chip'' 
devices has led to a stringent need of scaling standard machinery 
down to micro- and nano-scales. This reduction in size has raised a 
number of challenging issues, such as to endow small objects with the 
capacity to perform autonomous, directional motion 
\cite{Whitesides_2002,Paxton_review_2006,Lauga_2009,Howse_review_2010}. 
Towards this goal two main routes are pursued. The first one 
consists of designing artificial mechanical ``swimmers'' by mimicking 
the sophisticated locomotion strategies of natural micro-organisms 
such as \textit{E. Coli} or \textit{Spiroplasma} (see, e.g., Refs. 
\cite{Lauga_2009,Cheang_2010} and references therein). The second 
approach consists of transforming chemical free energy into mechanical 
work by employing phoretic mechanisms, i.e., motion induced by 
interfacial interactions \cite{Paxton_review_2006,Howse_review_2010}. 
In the following we focus on this latter approach.

Several proposals for such catalytic self-propellers have already 
been tested experimentally (see, e.g., Refs. 
\cite{Whitesides_2002,Paxton_2004,Sen_2005,Howse_2007,Leiderer_2008,
Sanchez_2009} and Refs. \cite{Paxton_review_2006,Howse_review_2010} 
for reviews). The underlying idea \cite{Whitesides_2002} is that an 
asymmetric decoration of the surface of a particle with a catalyst, 
which promotes an activated reaction in the surrounding liquid medium, 
leads to a non-uniform distribution of product molecules around the 
surface of the particle. This non-uniform distribution gives rise to 
particle motion through a variety of mechanisms 
\cite{Whitesides_2002,Paxton_2004,Howse_2007,Sanchez_2009}. When the 
size of the particle is decreased towards the micron scale or below, 
viscous and surface forces start to dominate and inertia-based 
mechanisms such as bubble ejection propulsion become ineffective 
\cite{Lauga_2009}. (However, alternative propulsion mechanisms based 
on bubble formation can remain active, as for the catalytically active 
tubes proposed in Ref. \cite{Sanchez_2009}.) If the product molecules 
remain dissolved in the surrounding liquid medium, concentration 
gradients develop along the surface of the particle. It has therefore 
been argued \cite{Paxton_2004,Golestanian_2005} that in such cases 
the motion of the catalyst-covered active particle is phoretic, i.e., 
it results from the interfacial interactions between the particle and 
the non-uniformly distributed product molecules generated by the 
chemical reaction. Recent experimental \cite{Paxton_2004,Paxton_2005,
Paxton_review_2006,Howse_2007,Leiderer_2008,Sanchez_2009,Boquet_2010}, 
simulation \cite{Kapral_2007,Lowen_2010}, and theoretical 
\cite{Golestanian_2005,Golestanian_2007,Prost_2009,Popescu_2009,
Popescu_2010,Boquet_2010,Lowen_2010} studies of such systems 
have contributed to a significantly improved understanding of this 
self-induced phoresis of isolated active particles.

One of the envisioned applications of such self-propellers is to use 
them as active carriers for colloidal transport and assembly 
\cite{Sen_2008,Sanchez_2009}. While in classic phoresis 
\cite{Derjaguin_1966,Anderson_1989} the concentration gradients are 
externally imposed and maintained, for self-phoresis the gradients 
are dynamically generated by the catalytic reaction occurring on 
parts of the particle surface. The interplay between these 
concentration gradients and the ensuing hydrodynamic flows not only 
powers the phoretic motion of such active particles, but also 
influences the effective interaction among each other, with nearby 
inert particles, or with bounding walls \cite{Golestanian_2007,
Popescu_2009}. Accordingly, the performance of such active particles 
as carriers is expected to depend strongly on these effective 
interactions. 

\begin{figure*}[!htb]
\begin{center}
\includegraphics[width = .28 \linewidth]{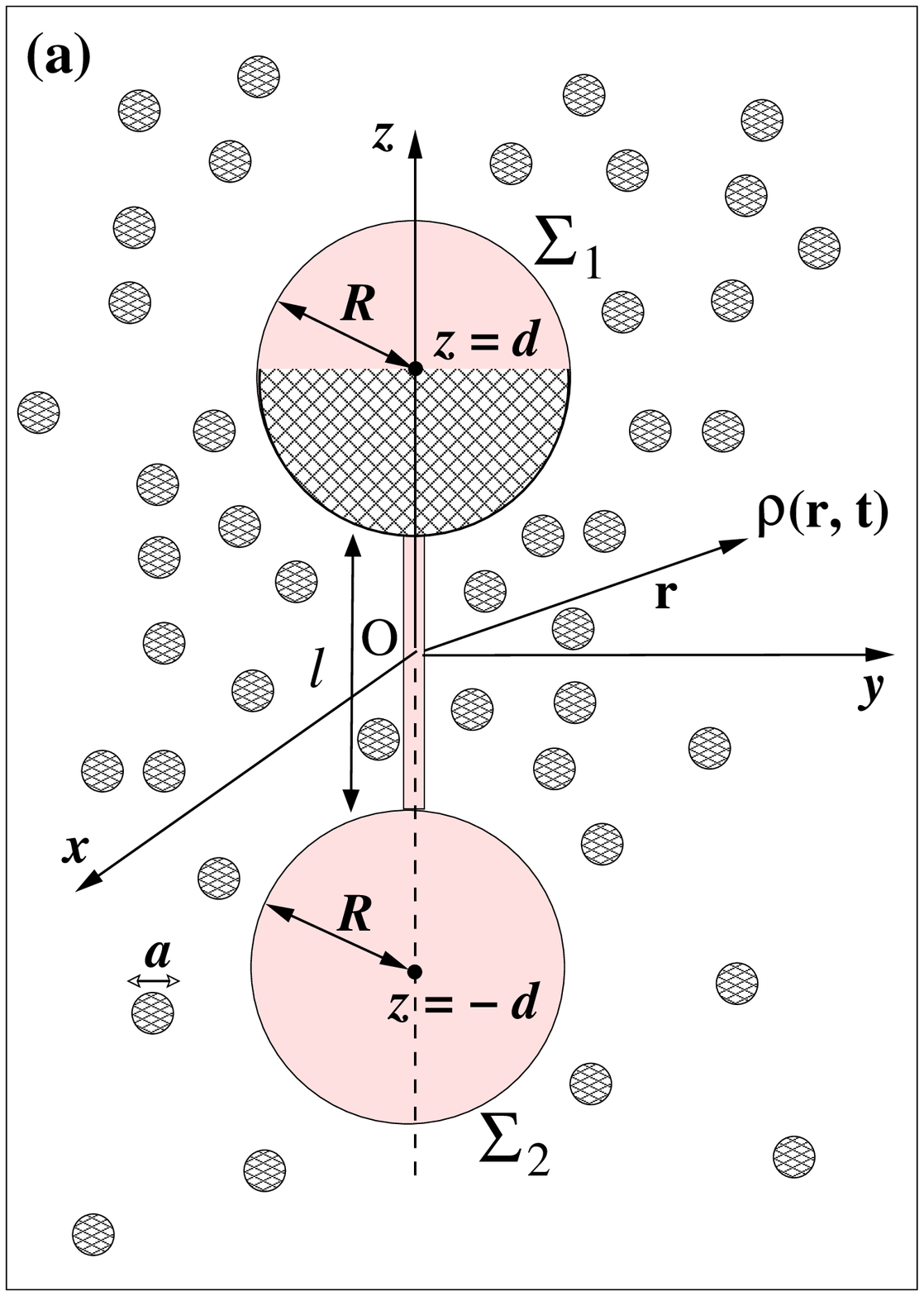}%
\hspace*{.02\linewidth}%
\includegraphics[width = .28 \linewidth]{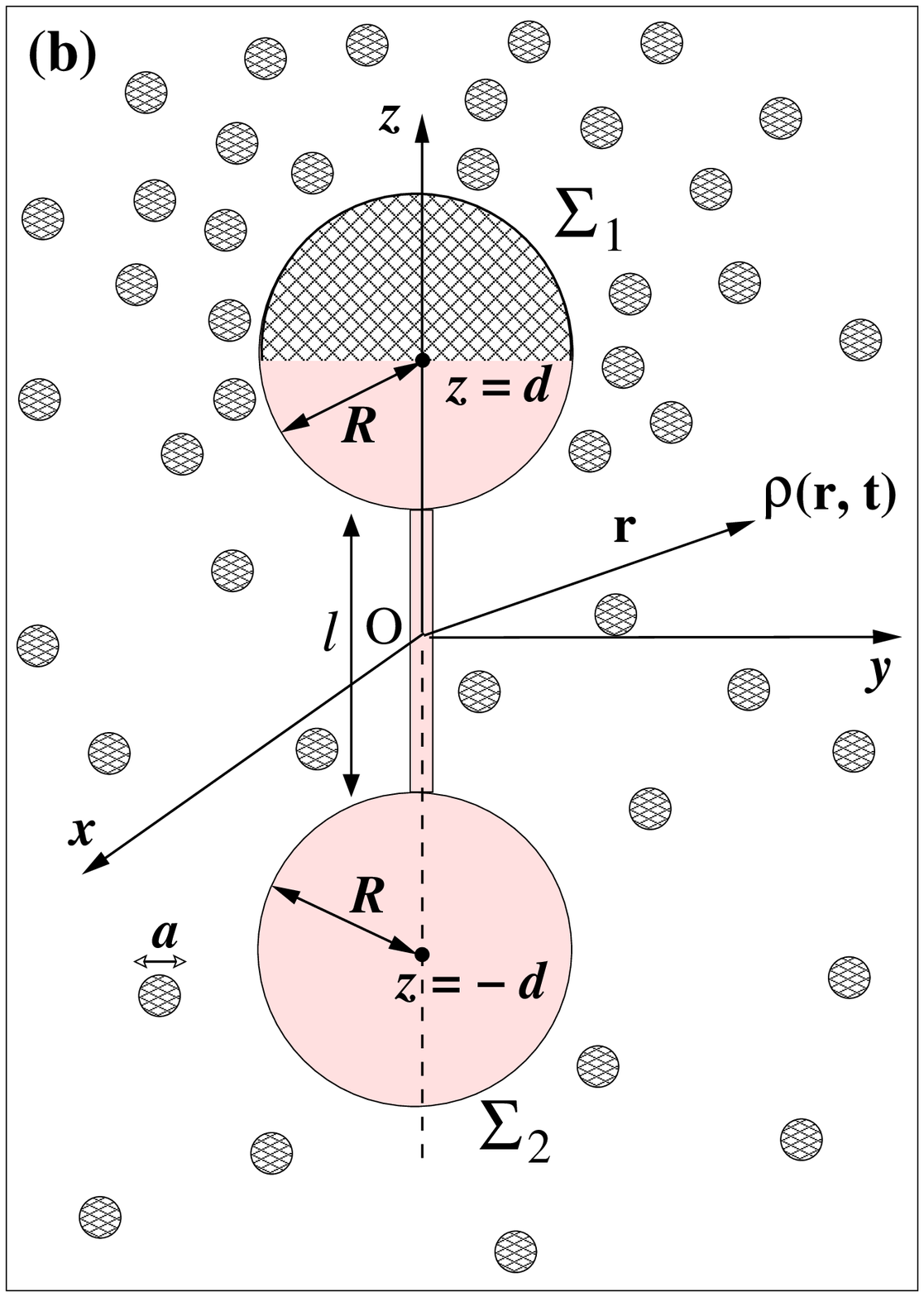}%
\hspace*{.02 \linewidth}%
\includegraphics[width = .28 \linewidth]{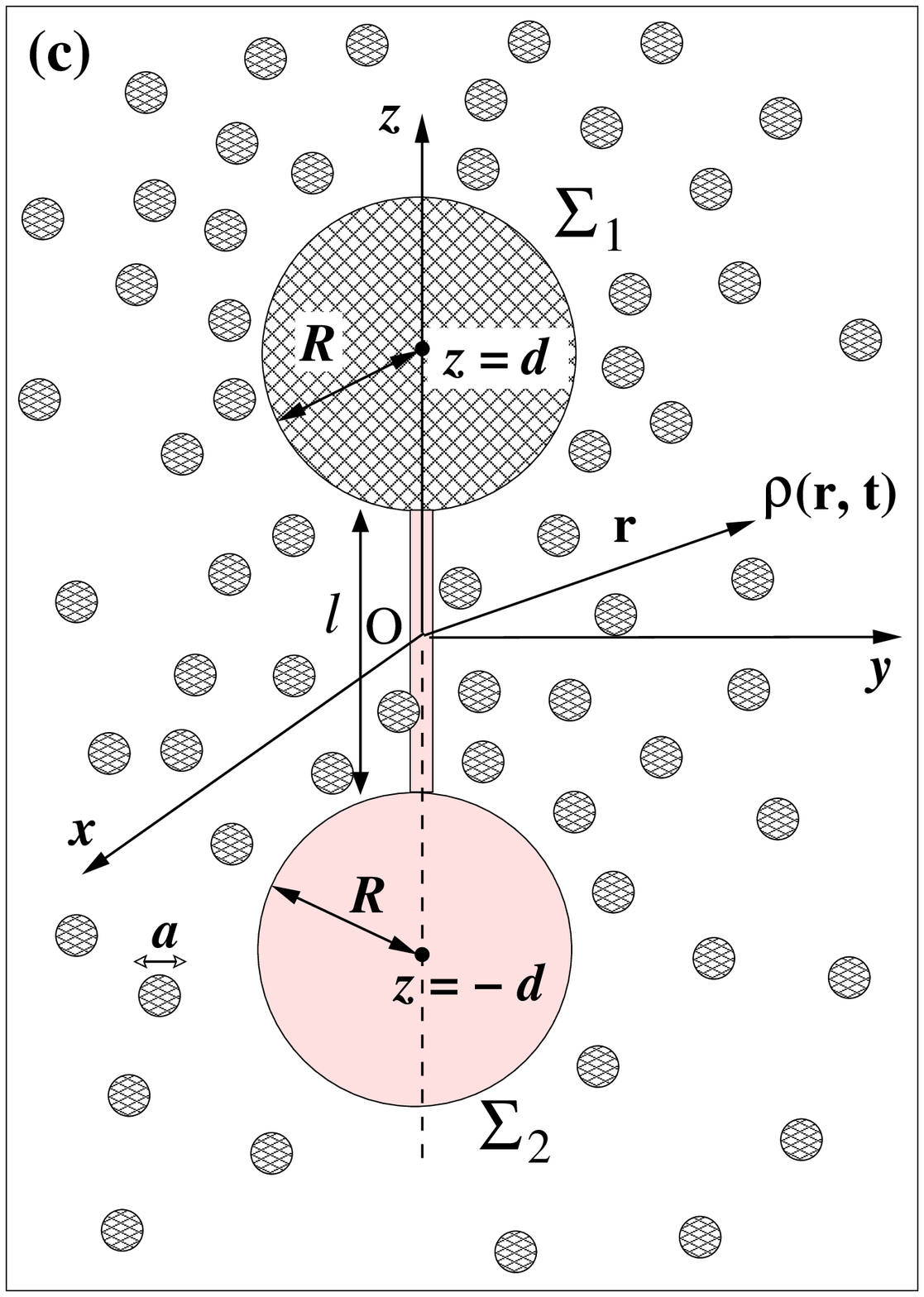}
\end{center}
\caption{
Carrier-cargo system consisting of a hard spherical 
carrier (top circle) of radius $R$ with 
its surface $\Sigma_1$ covered partially (a), (b) or completely 
(c) by a catalyst (hatched area) over a cap-like region. The carrier 
is linked via a thin rigid rod  of length $l = 2 (d - R)$ to a hard spherical, 
and inert cargo with surface $\Sigma_2$ and radius $R$. The product 
molecules (small hatched circles) with number density 
$\rho(\mathbf{r}, t)$ and of diameter $a$ are emerging from the 
solvent (not shown) via a chemical reaction.
}
\label{fig1}
\end{figure*}
Here we study a simple model for the transport of an inert particle, 
posing as a cargo, by an active diffusio-phoretic carrier 
\footnote{``Diffusio-phoresis'' is restricted here to 
phoresis in gradients of electrically neutral 
solutes (often also denoted as ``chemo-phoresis'').}. 
The model consists of two spherical particles, with 
the surface of one of them partially covered by a catalyst and being 
connected by a thin rigid rod. Similar to previous studies 
\cite{Golestanian_2005,Golestanian_2007,Popescu_2009,
Popescu_2010,Prost_2009}, we adopt the standard theory of 
diffusio-phoresis \cite{Derjaguin_1966,Anderson_1989} and derive, 
within the constraints of this approach \cite{Popescu_2009}, 
the velocity of the carrier-cargo composite for two distinct orientations 
of the link [see Figs. \ref{fig1}(a)-(b)]. 
The results show that the performance of the carrier, 
i.e., the resulting velocity of the composite system, strongly 
depends on the orientation of the link, which is a paradigmatic 
consequence of the self-phoresis mechanism of the motion. 
As an extreme case we study how a spherical particle, 
which is completely covered by catalyst and thus is unable to move 
on its own, can act as a carrier once it is attached to a cargo 
[see Fig.~\ref{fig1}(c)]. We note that the emergence of motion in this 
latter case can be intuitively understood by realizing that in the limit 
of a very short linkage this configuration is equivalent with the so-called 
dimer model of an active particle 
\cite{Kapral_2007}.

\textbf{\textit{Model.}}
\label{sec_model} 
The system we consider is shown and characterized in 
Fig.~\ref{fig1}. We assume a uniform distribution, with the areal 
density $\sigma$, of the catalyst over a hemisphere of the carrier 
particle. The infinitesimal thin rod enforces that there is no relative 
motion between the carrier and the cargo. Furthermore, the rod is 
aligned such that the carrier-cargo  composite possesses axial 
symmetry. The co-moving coordinate system has its $z$ direction along 
the axis of symmetry and the $xy$ plane is the midplane between the 
two spheres. If the carrier is partially covered by catalyst, 
there are two possible configurations with axial 
symmetry: the active area of the carrier points towards 
[Fig. \ref{fig1}(a)] or away from [Fig. \ref{fig1}(b)] the cargo. 
For this system the only possible motion is a translation along the 
$z$-direction. The purpose of these simplifications is to sift out 
the main phenomena of interest, i.e., the dependence 
of the self-phoretic velocity on the internal orientation of the 
carrier-cargo system and, for the configuration shown in 
Fig. \ref{fig1}(c) (which can be seen as a dimer model like in 
Ref. \cite{Kapral_2007}), the emergence of motion solely 
from linking together an otherwise motionless active 
particle with an inert cargo.

The catalyst promotes the chemical conversion of the solvent 
surrounding the composite system into product molecules. Here we 
focus on the particular case in which the chemical conversion 
$A \stackrel{\text{cat}}{\to} A' + B$ of a solvent molecule ($A$) 
leads to two product molecules ($A'$ and $B$) only, one with 
properties and size being similar to the solvent itself 
($A' \approx A$), the other one ($B$) being significantly different 
\footnote{It has been argued that this is approximately 
the case for the Pt catalyzed decomposition in aqueous solution of 
hydrogen peroxide ($A$ = H$_2$O$_2$) into water ( $A'$ = H$_2$O) 
and oxygen ($B$ = O$_2$) 
\cite{Golestanian_2005,Howse_2007,Howse_review_2010,
Popescu_2010}.
}. 
Accordingly, in the following only this latter one is 
denoted as a product molecule and plays the role of a solute in 
the solvent. The product molecules of diameter $a$ diffuse into the 
solvent characterized by a diffusion coefficient $D$. Thus the net 
result of the chemical conversion can be viewed approximately as 
the generation of a solute only by an ensemble of independent 
sources uniformly distributed over a part of the carrier surface. 
The reaction rate $\nu_B$ at such a catalytic site, i.e., the 
number of product molecules created at that site per unit time $t$, 
is assumed to be independent of time. The number density 
$\rho(\mathbf{r},t)$ of product molecules is considered to be so 
low that among themselves they behave like an ideal gas. However, 
there is an interaction potential between the product molecules 
and the moving particle, which, inter alia,  realizes the 
impermeability condition at the particle surface. The 
interactions between the product molecules and the solvent are 
accounted for effectively via the Stokes - Einstein 
expression $D = k_B T/(3 \pi \mu a)$ \cite{S-E_relation}, where 
$k_B$ is the Boltzmann constant, $T$ is the temperature, and 
$\mu$ is the viscosity of the solution (i.e., solvent plus solute).

The direct and the solvent mediated particle-solute interactions 
combined with the non-uniform distribution of the solute, created 
at the catalyst covered region, lead to an unbalanced osmotic 
pressure \textit{along} the surface of the particle which induces 
flow of the solvent 
\cite{Derjaguin_1966,Anderson_1989,Golestanian_2005,Ajdari_2006,
Prost_2009,Popescu_2009} and motion of the particle. For 
micrometer-size particles typical self-phoretic velocities are of 
the order of $\mu$m/s. Accordingly, both the Reynolds number 
$\mathrm{Re} \simeq \tilde \rho_{solv} \tilde V \tilde R /\mu$ 
(where $\tilde \rho_{solv}$ is the mass density of the solvent,  
$\tilde R$ the linear extension of the particle, and $\tilde V$ the 
velocity of the translational or rotational motion) and the Peclet 
number $\mathrm{Pe} \simeq \tilde V \tilde R /D$ are small 
\cite{Happel_book}. Thus the hydrodynamic description approximately 
reduces to the Stokes equations and the convection of the solute can 
be disregarded compared with its diffusive transport. 

\textbf{\textit{Phoretic slip.}}
\label{sec_phor_slip}
The effective interactions between the particles and the 
self-generated, asymmetric, non-uniform solute number density 
$\rho(\mathbf{r},t)$ around the  particles 
\cite{Derjaguin_1966,Anderson_1989,Golestanian_2007} have a typical 
range $\delta$, which is comparable with the solute diameter $a$, 
and induce flow of the solution relative to the particle. 
In steady 
state this hydrodynamic flow in the thin surface layer $\sim \delta$ 
is accounted for by a (phoretic) slip-velocity 
\cite{Derjaguin_1966,Anderson_1989},
\begin{equation}
\label{slip_vel}
\mathbf{v}_s (\mathbf{r}_p) = - b(\mathbf{r}_p) \nabla^\Sigma
\rho(\mathbf{r}_p)\,,\textrm{ for } \mathbf{r}_p \in \Sigma 
\,(= \Sigma_1 \cup \Sigma_2)\,,
\end{equation}
where $\mathbf{r}_p$ is a point $P$ on the surface $\Sigma = 
\Sigma_1 \cup \Sigma_2$ of the particles and $\nabla^\Sigma$ denotes 
the projection of the gradient operator onto the corresponding local 
tangential plane of the surface of the  particles 
\footnote{
Rigorously, Eq. (\ref{slip_vel}) should be interpreted as a condition 
on the outer edge $\Sigma_\delta$ of the surface layer. For the outer 
problem this can be replaced by $\Sigma$ because $\rho$ varies over 
length scales which are much larger than $\delta$.}. This 
serves as a boundary condition for the hydrodynamic flow in the outer 
region, i.e., outside the surface layer \cite{Anderson_1989,
Golestanian_2007,Popescu_2009}.
The mobility coefficient $b(\mathbf{r}_p)$ is determined by the 
aforementioned effective interaction potential and by the hydrodynamic 
boundary condition (stick or slip) on the surface of the particle, 
which is the \textit{inner edge} of the surface layer 
\cite{Anderson_1989,Ajdari_2006}.  For simplicity, here we take 
$b(\mathbf{r}_p)$ to be constant over the whole surface $\Sigma$. 
We note that $b \lessgtr 0$ correspond to repulsive 
and attractive effective interactions, 
respectively \cite{Popescu_2009,Anderson_1989}.

\textbf{\textit{Distribution of product molecules.}}
\label{sec_solute_dist}
In the limit of small Peclet numbers and neglecting any so-called 
polarization effects of the surface layer \cite{Anderson_1989}, the 
steady state distribution $\rho(\mathbf{r})$ of product molecules in 
the outer region around the two moving particles is governed, in the 
co-moving frame, by the diffusion equation
\begin{equation}
\label{diff_eq}
D \nabla^2 \rho(\mathbf{r}) = 0, \,\mathbf{r} \in
\text{outer region}\,.
\end{equation}
This is subject to the boundary conditions (BCs)
\begin{subequations}
\label{BC}
\begin{equation}
\label{farfield_BC}
\rho(|\mathbf{r}| \to \infty) = 0\,,
\end{equation}
\begin{equation}
\label{normalJ_BC}
- D \left.\left[\mathbf{\hat n} \cdot \nabla \rho(\mathbf{r})
\right]\right|_{\mathbf{r} \in \Sigma} = 
\nu_B \,\sigma\,\times
\begin{cases}
 & 1\,,~ {\mathbf{r}} \in \mathrm{~catalyst}\\
 & 0\,,\mathrm{~otherwise}
\end{cases}
\,.
\end{equation}
\end{subequations}
Equation (\ref{normalJ_BC}), where $\nu_B \,\sigma$ corresponds to the 
reaction rate per unit area and $\mathbf{\hat n}$ denotes the  outward 
direction normal to the surface, describes how the catalytic reaction 
at $\mathbf{r} \in \Sigma$ translates into a source of product 
particles $B$ \cite{Golestanian_2007,Popescu_2010}.

\textbf{\textit{Hydrodynamic flow outside the surface layer.}}
\label{sec_hydro_flow}
Because there are no forces acting on the solution beyond the 
surface layer, based on the assumption of low Reynolds numbers the 
hydrodynamic flow field $\mathbf{u}$ in the outer region is obtained 
as the solution of force free and incompressible Stokes equations:
\begin{equation}
\label{St_eq}
\nabla \cdot \mathbf{\hat \Pi} = 0\,,
~\nabla \cdot \mathbf{u} = 0\,.
\end{equation}
$\mathbf{\hat \Pi} := -p \mathbf{\hat I} + \mu \mathbf{\hat S}$ is 
the corresponding pressure tensor, where $p$ is the hydrostatic 
pressure and $\mathbf{\hat S}$ is the shear
stress tensor, i.e., $S_{\alpha \beta} =
\partial u_{\alpha}/\partial x_{\beta} +
\partial u_{\beta}/\partial x_{\alpha}$. 
In the comoving reference frame, attached to the carrier-cargo 
composite which is moving with $\mathbf{V} = V \mathbf{ \hat e}_z$ 
relative to the solution, which is quiescent far away, these 
equations are subject to the BCs of a flow velocity $-\mathbf{V}$ 
far away ($|\mathbf{r}| \to \infty$) from the composite and 
$\mathbf{v}_s (\mathbf{r}_p)$ at the outer edge of the surface layer. 
This corresponds to sticking at the surface of the particles plus a 
slip velocity $\mathbf{v}_s$ at the outer edge of the surface layer:
\begin{equation}
\label{BC_flow}
\left.\mathbf{u}\right|_{\Sigma} = \mathbf{v}_s\,,~~
\left.\mathbf{u}\right|_{|\mathbf{r}|\to \infty} = -\mathbf{V} \,.
\end{equation}
Equation (\ref{BC_flow}) reveals that the \textit{cargo} plays an 
\textit{active} role ($\mathbf{v}_s \neq 0$ on 
$\Sigma_2 \subset \Sigma$) despite being catalytically inert.

\textbf{\textit{Phoretic velocity.}}
\label{sec_Vphor}
After computing the hydrodynamic flow in the outer region, which 
depends parametrically on the translation of the two particles via 
the BCs in Eq. (\ref{BC_flow}), the phoretic velocity $\mathbf{V}$ is 
determined by the condition that the motion of the system composed of 
the particles plus their surface layers is force free 
\cite{Anderson_1989}. 

Equations (\ref{diff_eq}) and (\ref{BC}) as well as Eqs. (\ref{St_eq}) 
and (\ref{BC_flow}) are most conveniently solved in terms of bispherical 
coordinates 
$(\xi \in \mathbb{R},0 \leq \eta \leq \pi, 0 \leq \phi < 2 \pi)$ 
\cite{Reed_1976,Jeffery_1912,Jeffery_1926,Arfken_book}
\begin{equation}
\label{bispher_coord}
\lbrace x\,,y\,,z \rbrace = \varkappa
\lbrace  \sin\eta \cos\phi\,, \sin\eta \sin\phi\,, \sinh\xi \rbrace/C\,,
\end{equation}
where $C = \cosh\xi-\cos\eta$; 
the corresponding scale factors 
$h_\xi = |\partial \mathbf{r}/\partial \xi| 
= \left[(\partial x/\partial \xi)^2 + 
(\partial y/\partial \xi)^2 + (\partial z/\partial \xi)^2 \right]^{1/2}$, 
$h_\eta = |\partial \mathbf{r}/\partial \eta|$, and $h_\phi 
= |\partial \mathbf{r}/\partial \phi|$ are given by
\begin{equation}
\label{bispher_scale}
\lbrace h_\xi\,,h_\eta\,,h_\phi \rbrace = \varkappa
\lbrace  1\,, 1\,, \sin\eta \rbrace/C\,.
\end{equation}
The choice $\varkappa = \sqrt{d^2-R^2}$ ensures that the family of 
spheres $\xi  = const$ includes the ones, $\xi = \pm \xi_0 := \pm \,
\mathrm{arccosh}(d/R)$, corresponding to the surfaces $\Sigma_{1,2}$ 
of the carrier and the cargo, respectively. Noting that the 
intersection of an $\eta$ iso-surface with a $\xi$ iso-surface is a 
circle parallel to the $xy$ plane and that the equatorial circle 
parallel to the $xy$ plane on $\Sigma_1$ corresponds to $\eta_0 = 
\mathrm{arcctg}(R/\sqrt{d^2-R^2})$, the area covered by the catalyst 
is parametrized by $\xi = \xi_0 \,,~0 \leq \phi <  2 \pi$, and 
$\eta_0 \leq  \eta \leq \pi$ for the configuration in 
Fig. \ref{fig1}(a), $0 \leq \eta \leq \eta_0$ for the one in 
Fig. \ref{fig1}(b), and $0 \leq \eta \leq \pi$ for the one in 
Fig. \ref{fig1}(c).

In terms of bispherical coordinates the solution $\rho(\xi,\eta)$ 
of Eq. (\ref{diff_eq}), which does not depend on $\phi$ due to
the axial symmetry of the system, which is finite at $\eta = 0, \pi$ 
(i.e., on the $z$ axis), and which satisfies the BC in 
Eq. (\ref{farfield_BC}), can be written as 
\cite{Reed_1976,Jeffery_1912}
\begin{eqnarray}
\label{rho_bispher}
&&\rho(\xi,\omega:=\cos \eta) = 
\rho_0 (\cosh\xi - \omega)^{1/2} \,\sum_{n\geq0}\, P_n(\omega)
\nonumber\\
&& \times \left[A_n \sinh\left((n+1/2)\xi \right)
+ B_n \cosh\left((n+1/2)\xi\right)\right]\,,~~~
\end{eqnarray}
where $P_n$ denotes the Legendre polynomial of order 
$n$ and $\rho_0:= R \nu_B \sigma/D$ is a density scale 
chosen such that the system specific parameters are factored out 
from the boundary conditions [Eqs. \ref{BC}]. 
Noting that $\mathbf{\hat n} = (-,+) \mathbf{\hat e}_\xi 
= (+,-) 
\lbrace  
\sinh \xi \sin \eta \cos \phi \,, 
\sinh \xi \sin \eta \sin \phi\,, 
\cosh \xi \cos\eta-1 
\rbrace
/C$ 
on $\Sigma_{1,2}$ and 
expanding the right  hand side of  Eq. (\ref{normalJ_BC}) in 
a series of Legendre polynomials, the BCs in Eq. (\ref{normalJ_BC}) 
determine the coefficients $\lbrace A_n\,,B_n \rbrace$ as the 
solution of a system of linear equations. This determines 
$\rho(\xi,\eta)$ and the slip velocity follows from 
$\mathbf{v}_s = - 
[(b/h_\eta) \partial_\eta \rho(\xi,\eta)]_{\xi = \pm \xi_0} 
\,  \mathbf{\hat e}_\eta$ with 
$\mathbf{\hat e}_\eta = 
\lbrace  
(\cosh \xi \cos \eta - 1) \cos \phi \,, 
(\cosh \xi \cos \eta - 1)\sin \phi \,, 
- \sinh \xi \sin\eta
\rbrace
/C
$.

Due to the linearity of the Stokes equations the hydrodynamic flow 
field $\mathbf{u} =\mathbf{u}_I + \mathbf{u}_{II}$ 
[Eqs. (\ref{St_eq}) and (\ref{BC_flow})] is the superposition of 
the one corresponding to $\mathbf{u}_I({|\mathbf{r}|\to \infty}) 
= -\mathbf{V}$ with stick BCs on $\Sigma$ and the one corresponding 
to a quiescent flow far away, i.e., 
$\mathbf{u}_{II} (|\mathbf{r}| \to \infty$ = 0 with slip BCs 
$\mathbf{v}_s$  on $\Sigma$. The solution for the 
first problem is known \cite{Jeffery_1926} and corresponds to a 
hydrodynamic force on the composite 
$\mathbf{F}_I = - 12 \pi \mu R \lambda \mathbf{V}$, where 
\begin{eqnarray}
\label{lambda}
&&\lambda = \dfrac{4}{3} \sinh \xi_0 \sum_{n \geq 1} 
\dfrac{n (n+1)}{(2n-1)(2n+3)} \nonumber\\
&&\times \left[1-\dfrac{4 \sinh^2 (n+1/2)\xi_0 
- (2 n +1)^2 \sinh^2 \xi_0}{2 \sinh (2 n+1) \xi_0  
+ (2 n+1)\sinh 2 \xi_0}\right] \,.
\end{eqnarray} 
The second solution $\mathbf{u}_{II}$ for the flow field is most 
conveniently expressed in terms of the Stokes stream function $\Psi$ 
for axisymmetric flows [which in cylindrical coordinates 
$\lbrace \tilde r = \sqrt{x^2 + y^2}, \phi, z \rbrace$ is defined by 
$(u_{\tilde r}\,,u_{z}) :=  
(- \partial_{z} \Psi \,,\partial_{\tilde r} \Psi)/\tilde r $]. 
In terms of
bispherical coordinates $\Psi(\xi,\omega:= \cos\eta)$ is given by
\cite{Jeffery_1926}
\begin{equation}
\label{stream_func}
\Psi(\xi,\omega) = V_0 R^2 \sum_{n\geq 0}
\dfrac{W_n(\xi) \, C^{(-1/2)}_{n+1}(\omega)}{(\cosh\xi-\omega)^{3/2}}\,,
\end{equation}
where
\begin{eqnarray}
&& W_n(\xi) = a_n \cosh\left((n-1/2) \xi\right) +  
b_n \sinh\left((n-1/2) \xi \right)~~ \nonumber\\
&& + c_n \cosh\left((n+3/2)\xi\right) + d_n \sinh\left((n+3/2)\xi\right)\,,
\end{eqnarray}
$C^{(-1/2)}_{n}(\omega) = [P_{n-2}(\omega) - P_{n}(\omega)]/(2 n-1)$ 
\cite{Jeffery_1926} is formally denoted as the Gegenbauer 
polynomial of degree $n$ and order $-1/2$ \cite{Brenner_1961}, and 
the characteristic velocity 
$V_0 = b \rho_0/R$ is chosen such that the system dependent parameters 
are factored out from the boundary conditions in Eq. (\ref{BC_flow}). 
Because there are no flow sources or sinks along the $z$-axis, we require 
the stream function to vanish there \cite{Happel_book}, i.e., 
$\Psi(\xi,\omega = \pm 1) = 0$. 
Since $C^{(-1/2)}_{m}(\omega = \pm 1) = 0$ for $m \geq 2$, but as a 
linear function $C^{(-1/2)}_{1}(\omega)$ cannot be zero for both 
$\omega = \pm 1$, this constraint implies that the term $n = 0$ is 
removed from the series representation in Eq. (\ref{stream_func}).

In terms of the stream function, the boundary condition of a vanishing 
flow velocity along the direction normal to the surfaces $\Sigma_{1,2}$ 
(corresponding to impenetrability) leads to the requirement 
$\left[\partial\Psi(\xi,\omega = \cos \eta)/\partial\eta 
= 0\right]_{\pm \xi_0} = 0$ \cite{Happel_book}, which can be shown to be 
equivalent to the simpler form $\Psi(\pm \xi_0,\omega) = 0$ 
\cite{Happel_book,Brenner_1961}.
The flow velocity along the slip direction $\eta$ is given in terms of 
the stream function by $u_\eta 
= - C^2\,(\varkappa^2 \,\sin\eta)^{-1} \partial_\xi \Psi$ 
\cite{Happel_book} which, according to Eq. (\ref{BC_flow}), 
on the surfaces $\Sigma_{1,2}$ should be equal to the phoretic slip 
$\mathbf{v}_s$. Since for $n \neq m$ and $n,m \geq 2$ the 
polynomials $C^{(-1/2)}_{n,m}$ are orthogonal, 
$\int_{-1}^{1} d \omega (1-\omega^2)^{-1} 
C^{(-1/2)}_{n}(\omega) C^{(-1/2)}_{m}(\omega) = 0$, the 
coefficients $\lbrace a_n, \dots, d_n \rbrace$ can be determined
from these boundary conditions on $\Psi$ in terms of 
$\lbrace A_n, B_n \rbrace$ by solving a system of linear equations 
\footnote{In practice, the infinite system has to be 
truncated at a certain order $n = N_{max}$. For our system it turns 
out that both $A_n$ and $B_n$ vanish rapidly upon increasing $n$. 
In all cases studied, at $N_{max} = 100$ these coefficients are 
smaller than $10^{-20}$, compared with values of the order of 
unity for $n = 0$. This behavior is sufficient for the convergence 
of the velocity expression in Eq.~(\ref{velocity}).}. 
The hydrodynamic force acting in the positive $z$-direction on the 
composite due to this second flow $\mathbf{u}_{II}$ is given by 
$F_{II} = (2^{5/2} \pi \mu V_0 R^2/\varkappa) 
\sum_{n \geq 1} (a_n + c_n)$ 
\cite{Jeffery_1926}.

\begin{figure*}[!htb]
\begin{center}
\includegraphics[width = .33 \linewidth]{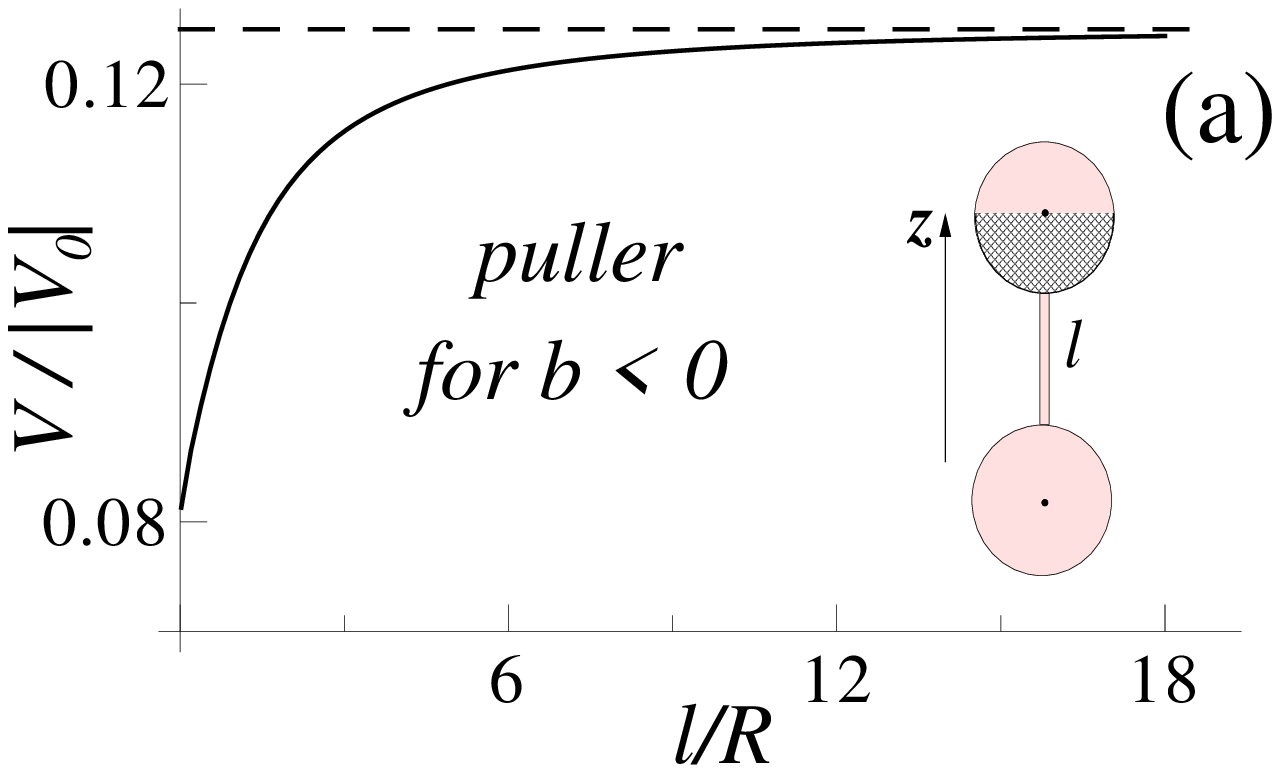}%
\hspace*{.015\linewidth}%
\includegraphics[width = .32 \linewidth]{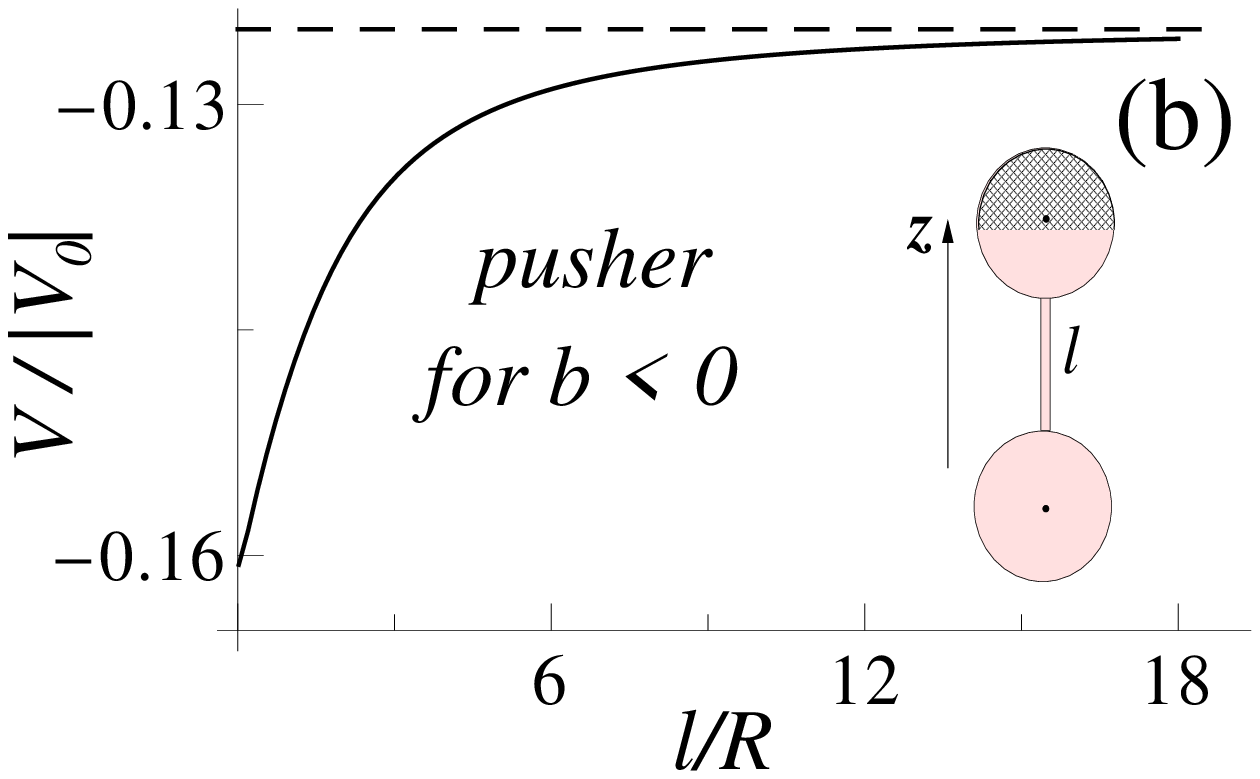}%
\hspace*{.015\linewidth}%
\includegraphics[width = .32 \linewidth]{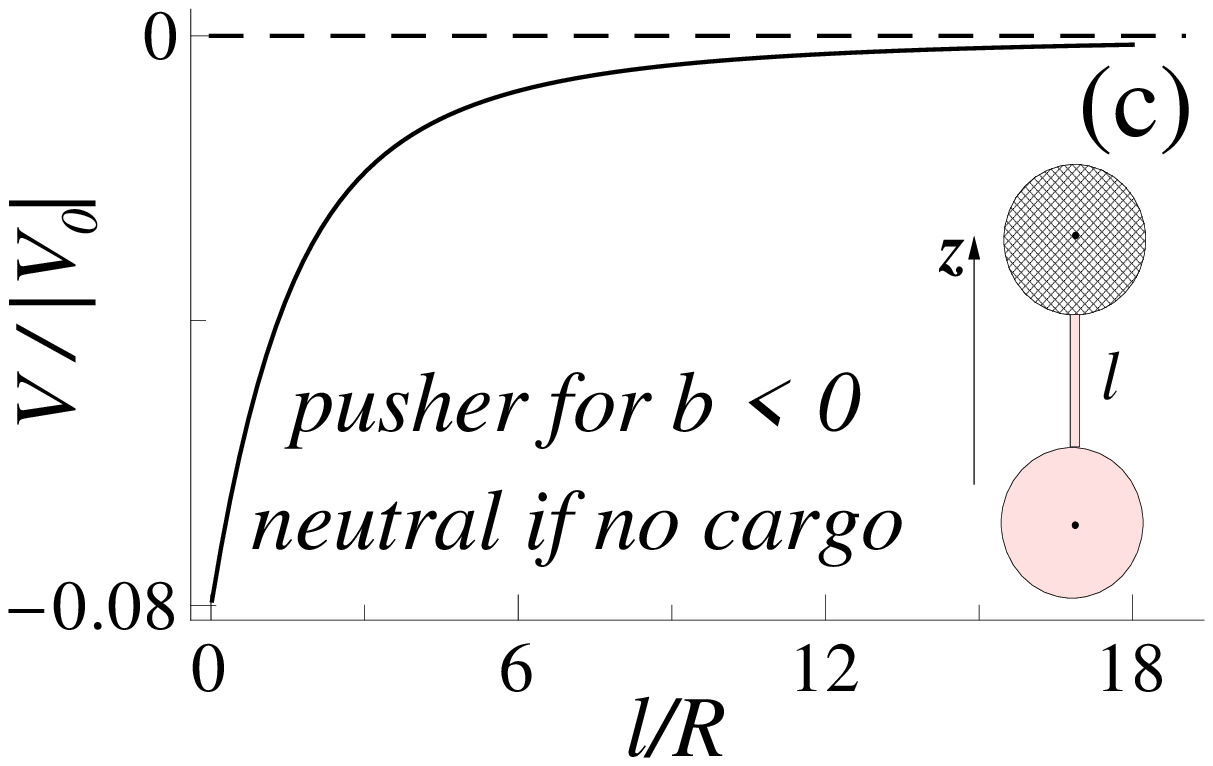}
\end{center}
\caption{
The velocity $V$ along the $z$-direction (indicated in the insets) 
in units of $|V_0|$ as a function of the scaled length 
$0.02 \leq l/R \leq 18$ of the linkage for the three 
configurations shown in the corresponding insets and for the case that 
the effective interaction between the product molecules and the two 
particles is repulsive ($b < 0$). The dashed lines in (a) and (b) 
correspond to the values $\pm |V_0|/8$, respectively, which the 
velocity of the cargo-carrier composite seems to attain as the cargo 
is linked at a larger distance from the carrier.
}
\label{fig2}
\end{figure*}
Requiring that the total force $\mathbf{F}_I + \mathbf{F}_{II}$ 
acting on the composite vanishes \cite{Derjaguin_1966,Anderson_1989} 
thus leads to the following expression for the velocity:
\begin{equation}
\label{velocity}
V/V_0 = \,\dfrac{\sqrt{2}}{3 \lambda \sqrt{(d/R)^2-1}} \,
\sum_{n \geq 1} (a_n + c_n)\,.
\end{equation}

\textbf{\textit{Discussion.}}
\label{sec_Disc}
In Fig. \ref{fig2} we show the velocity of the carrier-cargo 
system as a function of the scaled length $l/R = 2 \,(d/R - 1)$ 
of the linkage in the range $0.02 \leq l/R \leq 18$ for the three 
configurations considered in Fig. \ref{fig1} in the case of 
\textit{repulsive} effective interactions between the product molecules 
and the material forming the carrier-cargo system, i.e., for $b < 0$ 
and $V_0 < 0$. 
(The calculations can easily be extended to $l/R < 0.02$, but: 
(i) the implicit assumptions of a homogeneous solvent and point-like 
product molecules are expected to break down for very small 
separations, which renders Eq. (\ref{velocity}) to be inapplicable 
for too small values of $l/R$, and (ii) the limiting configuration 
$l/R = 0$ corresponding to carrier and cargo being in contact merely 
reduces to the case of self-diffusiophoresis of a single particle 
with a complex shape.)

For the choice $b < 0$ the carrier alone moves against the gradient 
of product molecule along its surface. Therefore the configurations 
in Figs. \ref{fig1} (a) and (b) correspond to a pulling and a 
pushing carrier, respectively. Focusing first on the configurations 
in Figs. \ref{fig2} (a) and 
(b), we note that the pusher or puller character is maintained at 
all separations $l$, i.e., the velocity of the composite is always 
along the positive and negative $z$ direction in the cases (a) 
and (b), respectively. In all cases, for $l/R \lesssim 4$ the 
velocity shows a significant dependence on the separation $l$, which 
clearly indicates that the cargo plays actually an active role in 
the resulting motion of the composite. In both cases at 
large separations $l$ the velocity approaches limiting values 
seemingly equal to $\pm |V_0|/8$, respectively, which is  $1/2$ of 
the velocity at which the active carrier half-covered by catalyst 
would move if isolated \cite{Popescu_2010}. However, the pusher 
slows down upon increasing $l$ whereas the puller speeds up. 
Therefore, for any length $l$ of the linkage the pusher 
configuration exhibits a faster motion than the puller one. These 
findings can be rationalized intuitively as follows. For example, 
in the case (a) the generation of product molecules by the carrier 
induces also a concentration gradient around the cargo surface 
pointing in the direction opposite to the one on the carrier 
surface. Consequently, the cargo behaves also like an active 
particle and induces a hydrodynamic flow around its surface in the 
direction opposite to the one around the carrier, which thus leads 
to a smaller net hydrodynamic flow and accordingly to a slower 
motion of the composite. 

Any other orientation of the carrier-cargo link breaks the axial 
symmetry, inducing a rotation of the composite in addition to its 
translation; this leads to a reduction in the translational velocity. 
Therefore the pusher configuration [Fig. \ref{fig1} (b)] is the 
optimal (i.e., maximal velocity) configuration if the effective 
interaction between the product molecules and the carrier-cargo 
material is repulsive. On the other hand, if this effective 
interaction is attractive ($b > 0$) the same intuitive arguments 
as above lead to the conclusion that the puller and pusher 
characters are interchanged between the configurations shown in 
Fig. \ref{fig1}(a) and (b), while the $l$-dependence of the 
velocity changes signs. Therefore, for attractive effective 
interactions the puller configuration [Fig. \ref{fig1} (b)] is 
the optimal one.

The motion of active particles is known to be affected by 
thermal fluctuations, and thus its directionality only holds 
over time scales smaller than the characteristic rotational 
diffusion time, crossing over to Brownian motion at longer time 
scales \cite{Golestanian_2007,Golestanian_2009}. Inspecting 
the cases of the carrier-cargo configurations shown in Figs. 
\ref{fig1} (a) and (b), one realizes that in (a) the cargo has 
a stronger influence than in (b) on the stability of the composite 
with respect to such fluctuations of the direction of motion. 
For $b<0$, in the case of the puller [Fig. \ref{fig1} (a)] a 
sudden change in direction producing a tilt of the 
symmetry axis towards the left relative to the O$z$ axis breaks 
the axial symmetry of the product concentration profile, and 
larger concentrations occur to the right of the symmetry axis. 
Consequently, right after the change in direction the cargo is 
driven through a region where the concentration of the product 
particles is larger on its right half (relative to the 
cargo-carrier axis) and thus induces a torque on the cargo and 
a rotation of the symmetry axis back towards the O$z$ direction. 
Therefore, in the puller 
configuration the cargo tends to stabilize the directionality of 
motion against thermal fluctuations. On the other hand, in the 
pusher configuration [Fig. \ref{fig1} (b)] the cargo is further 
away from the source of product particles. Therefore it is 
basically insensitive to such changes in the symmetry of the 
product concentration distribution as it moves away from it 
and consequently no significant restoring torques occur.

Configuration (c) reveals the rather peculiar consequence 
of self-phoretic propulsion that a carrier, which in isolation would 
be motionless, is activated by being linked to an otherwise inert 
cargo. This occurs because the linkage provides the anisotropy 
needed for self-phoresis to become operational 
\cite{Paxton_2004,Golestanian_2005}. For short links, the velocity 
of such a composite is significant. According to Fig. \ref{fig2}, 
for $l/R \lesssim 1$, $|V|/|V_0|$ is approximately 1/3 
of the largest possible velocity $|V_0|/4$ of an isolated spherical 
active carrier. For repulsive (attractive) effective 
interactions the inert cargo turns the neutral carrier into a pusher 
(puller).  We note that for lengths $l/R \ll 1$ the carrier-cargo 
system in configuration (c) becomes 
similar to the geometry of the so-called dimer model of an active 
colloid which, for an $A \to B$ catalytic reaction, has been proposed 
and investigated with Molecular Dynamics simulations 
\cite{Kapral_2007}.

The diffusiophoresis mechanism and the geometry of the model have 
been chosen with the intention to provide clear examples which 
reveal and underscore the complexity of such active carrier - cargo 
systems. We note that the main influence of the inert cargo 
stems from the fact that the nearby chemically active colloid gives 
rise to concentration gradients, and thus to phoretic slip velocity, 
along its surface, too. This is very different from the cases discussed 
in Ref. \cite{swimmers_2008}, where hydrodynamic interactions between 
dimer swimmers are "rectifying" the otherwise reciprocal movement of 
each of two dimers, and thus are leading to the emergence of collective 
and relative motion. We anticipate a rather rich behavior to emerge as 
various model constraints are relaxed, such as: (i) Different effective 
interactions between the product molecules and the cargo and carrier 
material, i.e., different mobility factors $b$ (of potentially 
opposite sign) on the carrier and cargo surfaces, will lead to a 
complex dependence of the velocity on the orientation of the linkage.
(For example, in the case shown in Fig. \ref{fig2}(a) with $b_{1,2}$ 
of opposite sign the velocity can change sign as $|b_2|$ is 
increased.) (ii) In the case of charged active particles and charged 
reaction products, the motion will be determined by the interplay 
of self-diffusiophoresis and self-electrophoresis, which amplifies 
the roles played by the orientation of the linkage and by the 
surface properties of the cargo and carrier. (iii) A convex
or concave shape of the surface of the cargo facing the carrier 
side leads to a possible amplification or reduction, respectively, 
of the concentration gradients around the carrier and the cargo. 
This list can be extended, but it already now shows that employing 
active particles as carriers allows an exceptional flexibility in 
the design of cargo-carrier systems, which should be very beneficial 
for potential applications.


\end{document}